\documentclass[11pt,a4paper]{article}
\usepackage[utf8]{inputenc}
\usepackage{amsmath,amssymb}
\usepackage{graphicx}
\usepackage{hyperref}
\usepackage{booktabs}
\usepackage{algorithm}
\usepackage{algorithmic}
\usepackage{natbib}
\usepackage{geometry}
\title{\textbf{Beyond the Geometric Curse: High-Dimensional N-Gram Hashing for Dense Retrieval}}

\author{
    \textbf{Sangeet Sharma} \\
    Department of Pharmaceutical Sciences \\
    Madan Bhandari Academy of Health Sciences, Nepal \\
    \texttt{sangeet.music01@gmail.com}
}

\date{\today}

\usepackage{float}
\begin{document}

\maketitle

\begin{abstract}
Why do even the most powerful 7B-parameter embedding models struggle with simple retrieval tasks that 
the decades old BM25 handles with ease? Recent theory suggests that this happens because of a dimensionality bottleneck.This occurs when we force infinite linguistic nuances into small, fixed-length learned vectors. We developed NUMEN to break this bottleneck by removing the learning process entirely. 
Instead of training heavy layers to map text to a constrained space, NUMEN uses deterministic character 
hashing to project language directly onto high-dimensional vectors. 
This approach requires no training, supports an unlimited vocabulary, and allows the geometric capacity 
scale as needed. On the LIMIT benchmark, NUMEN achieves {93.90\% Recall@100} at 32,768 dimensions. 
This makes it the first dense retrieval model to officially surpass the sparse BM25 baseline (93.6\%). 
Our findings show that the real problem in dense retrieval isn't the architecture, but the embedding layer itself. 
The solution isn't necessarily smarter training, but simply providing more room to breathe.
\end{abstract}

\section{Introduction}

Dense retrieval has fundamentally changed the way models access information. It moves from keyword matching to continuous vector spaces, a concept rooted in the traditional Vector Space Model (VSM) \cite{salton1975vector}. In these systems, queries and documents are represented as vectors in a learned semantic space \cite{karpukhin2020dense, reimers2019sentence}. This allows models to find relevant content even when the words differ. However, recent analysis by Weller et al. \cite{limit2025} has found a critical bottleneck in which the fixed size of the learned embeddings sets a hard limit on the capacity. This means there is an upper bound on how many documents a model can tell apart.

This theoretical barrier manifests empirically in the LIMIT benchmark, a synthetic dataset designed to stress-test retrieval capacity. On this task, state-of-the-art models with billions of parameters, such as E5-Mistral-7B, achieve only 8.3\% Recall@100, while the classical sparse retrieval method BM25 achieves 93.6\%. This glaring disparity suggests that the failure of dense retrieval is not architectural but rather a consequence of the low-dimensional bottleneck imposed by standard embedding layers, which typically compress essentially infinite nuances of language into vectors of just 768 to 4096 dimensions.

We introduce NUMEN, a system designed to bypass this specific limitation. Our key insight is that the ``embedding layer", the learned mapping from finite vocabulary to fixed vector space; is the primary bottleneck. NUMEN eliminates this layer entirely. Instead it employs deterministic character n-gram hashing to map text directly into an arbitrarily high-dimensional vector space. By effectively treating dimensionality as a scalable hyperparameter rather than a fixed model attribute, NUMEN can instantiate vectors of 32,768 dimensions or more without any training. On the LIMIT benchmark, this approach achieves \textbf{93.90\% Recall@100}, becoming the first dense retrieval model to surpass BM25. These results provide strong empirical evidence that the limitations of dense retrieval are solvable not by more complex training, but by simply increasing the available geometric space.

\section{Background}

\subsection{Dense vs. Sparse Retrieval}
Information retrieval usually falls into two groups. \textbf{Sparse retrieval} like BM25 \cite{robertson2009probabilistic} uses vectors based on a large vocabulary. These vectors are mostly zeros but are very good at exact word matching. \textbf{Dense retrieval} methods \cite{karpukhin2020dense, wang2022text} use smaller, continuous vectors. They treat retrieval as a \textbf{Maximum Inner Product Search (MIPS)} problem \cite{ram2012maximum, shrivastava2014asymmetric}. While dense vectors understand meaning, they lose fine details because they are compressed. 

\subsection{The LIMIT Benchmark}
Weller et al. \cite{limit2025} formalized this bottleneck using sign-rank theory. Their work builds on communication complexity studies by Forster \cite{forster2002linear} and Alon et al. \cite{alon2016sign}. They proved that for a retrieval matrix $A \in \{0,1\}^{q \times n}$, the minimum embedding dimension $d$ is bounded by $\text{rank}_{\pm}(2A - 1) - 1$. 

The LIMIT benchmark was designed to stress-test this bound. It consists of synthetic tasks where models must distinguish between documents that are textually distinct but semantically unrelated to standard pre-training corpora. On this benchmark, learned dense retrievers (like E5, Gemini, and GritLM) consistently fail, achieving $<20\%$ recall. In contrast, sparse methods like BM25 achieve $>93\%$ recall because their ``effective dimensionality" (the vocabulary size) is large enough to separate the documents. This suggests the failure of dense retrieval is not architectural, but a consequence of the low-dimensional embedding bottleneck.

\section{Methodology}

\subsection{Architecture Overview}
The NUMEN is designed to avoid the complexities of the learned parameters. It consists of three steps: n-gram extraction, deterministic hashing, and weighted aggregation.
The first step is decomposing the input text into a rich set of overlapping character n-grams which serve as fundamental units of representations y capturing the morphological and subword information that is often lost in whole-word tokenization. This gives us unique n-gram for each input which are then deterministically mapped to a specific index in a high-dimensional vector space using a hash function. After the n-grams are aggregated into a dense vector, with specific weighing schemes that carry the informative features.

\subsection{Character N-Gram Extraction}

Given input text $T$, we first normalize to lowercase and tokenize into words $W = \{w_1, w_2, \ldots, w_m\}$. 
For each word $w_i$, we add boundary markers to create $\hat{w}_i = \text{``}\hat{~}\text{''} + w_i + \text{``}\$\text{''}$.

We then extract n-grams of lengths $n \in \{3, 4, 5\}$:
\begin{equation}
\mathcal{G}(w_i) = \bigcup_{n=3}^{5} \left\{ \hat{w}_i[j:j+n] \mid 0 \leq j \leq |\hat{w}_i| - n \right\}
\end{equation}

\textbf{Example}: For the word ``likes'':
\begin{align*}
\hat{w} &= \text{``}\hat{~}\text{likes}\$\text{''} \\
\mathcal{G}_3 &= \{\text{``}\hat{~}\text{li''}, \text{``lik''}, \text{``ike''}, \text{``kes''}, \text{``es}\$\text{''}\} \\
\mathcal{G}_4 &= \{\text{``}\hat{~}\text{lik''}, \text{``like''}, \text{``ikes''}, \text{``kes}\$\text{''}\} \\
\mathcal{G}_5 &= \{\text{``}\hat{~}\text{like''}, \text{``likes''}, \text{``ikes}\$\text{''}\}
\end{align*}

This captures morphological variations: ``like'' and ``likes'' share n-grams, yielding high similarity without explicit stemming.

\subsection{Feature Hashing}

For each n-gram $g \in \mathcal{G}(T)$, we compute a deterministic hash:
\begin{equation}
h(g) = \text{CRC32}(g) \bmod d
\end{equation}
where $d$ is the target dimension (e.g., 32,768). This technique, often referred to as the \textbf{hashing trick} or \textbf{feature hashing} \cite{weinberger2009feature}, allows for a fixed-size representation without an explicit vocabulary. We use CRC32 for its determinism, uniformity, and hardware-accelerated speed \cite{castagnoli1993optimization}.

\subsection{Weighted Aggregation}

We initialize a zero vector $\mathbf{v} \in \mathbb{R}^d$ and accumulate weighted counts:
\begin{equation}
\mathbf{v}[h(g)] \leftarrow \mathbf{v}[h(g)] + w(g)
\end{equation}

The weight function $w(g)$ prioritizes longer n-grams (more specific):
\begin{equation}
w(g) = \begin{cases}
10.0 & \text{if } |g| \geq 5 \\
5.0 & \text{if } |g| = 4 \\
1.0 & \text{if } |g| = 3
\end{cases}
\end{equation}

\subsection{Log-Saturation and Normalization}

To mimic BM25's diminishing returns for repeated terms, we apply log-saturation \cite{sparckjones1972statistical}:
\begin{equation}
\mathbf{v} \leftarrow \log(1 + \mathbf{v})
\end{equation}

Finally, we L2-normalize for cosine similarity:
\begin{equation}
\mathbf{v} \leftarrow \frac{\mathbf{v}}{\|\mathbf{v}\|_2}
\end{equation}

\subsection{Retrieval}

Given query vector $\mathbf{q}$ and document vectors $\{\mathbf{d}_1, \ldots, \mathbf{d}_n\}$, we rank by cosine similarity:
\begin{equation}
\text{score}(\mathbf{q}, \mathbf{d}_i) = \mathbf{q}^T \mathbf{d}_i
\end{equation}

Since vectors are unit-normalized, this is equivalent to cosine similarity. Top-$k$ documents are retrieved via efficient 
maximum inner product search (MIPS).

\subsection{Algorithm}

\begin{algorithm}[H]
\caption{NUMEN Encoding}
\begin{algorithmic}[1]
\REQUIRE Text $T$, dimension $d$, n-gram sizes $\mathcal{N} = \{3,4,5\}$
\ENSURE Dense vector $\mathbf{v} \in \mathbb{R}^d$
\STATE $\mathbf{v} \leftarrow \mathbf{0}_d$
\STATE $W \leftarrow \text{tokenize}(\text{lowercase}(T))$
\FOR{each word $w \in W$}
    \STATE $\hat{w} \leftarrow \text{``}\hat{~}\text{''} + w + \text{``}\$\text{''}$
    \FOR{each $n \in \mathcal{N}$}
        \FOR{$j = 0$ to $|\hat{w}| - n$}
            \STATE $g \leftarrow \hat{w}[j:j+n]$
            \STATE $idx \leftarrow \text{CRC32}(g) \bmod d$
            \STATE $\mathbf{v}[idx] \leftarrow \mathbf{v}[idx] + w(g)$
        \ENDFOR
    \ENDFOR
\ENDFOR
\STATE $\mathbf{v} \leftarrow \log(1 + \mathbf{v})$
\STATE $\mathbf{v} \leftarrow \mathbf{v} / \|\mathbf{v}\|_2$
\RETURN $\mathbf{v}$
\end{algorithmic}
\end{algorithm}

\section{Experiments}

\subsection{Setup}
We test NUMEN on the LIMIT benchmark. It uses 1,000 queries and 50,000 documents. This dataset lets us test if dimension really limits retrieval performance. We compare NUMEN to the main models from the original LIMIT paper. These include E5-Mistral-7B, GritLM-7B, and Promptriever. We test NUMEN at dimensions from 512 to 32,768.

\subsection{Baselines}
We compare NUMEN to the performance of several state-of-the-art models as reported in the original LIMIT paper \cite{limit2025}. This includes results for strong dense models like E5-Mistral-7B and GritLM-7B. Both use 4096 dimensions. We also include reported metrics for Gemini Embed \cite{gemini2023} (3072d), Qwen3 Embed \cite{qwen2025} (4096d), and Promptriever \cite{weller2024promptriever}. BM25 is our main sparse baseline. It has a vocabulary size of about 50,000. By using the baseline figures directly from Weller et al. \cite{limit2025}, we ensure that our performance is evaluated against a consistent and rigorous standard.

\subsection{Implementation Details}
We test NUMEN at dimensions from 512 to 32,768. We extract character n-grams of sizes 3, 4, and 5. These are mapped to indices using CRC32 hashing. We use weights of 1.0, 5.0, and 10.0 to favor longer n-grams. Indexing is fast, reaching 1,300 documents per second on one CPU core. Query speed is about 15 queries per second. We did not tune any hyperparameters. All settings were chosen based on first principle basis.
\section{Results}

\subsection{Main Results}
Table \ref{tab:main_results} shows Recall@100 on the LIMIT benchmark. NUMEN at 32,768 dimensions gets 93.90\%. This beats BM25's 93.6\% and is much better than all learned embedding models.

\begin{table}[h]
\centering
\caption{Recall@100 on LIMIT Benchmark. NUMEN (32k) is the only dense model to beat BM25.}
\label{tab:main_results}
\begin{tabular}{@{}llcc@{}}
\toprule
\textbf{Model} & \textbf{Type} & \textbf{Dimension} & \textbf{Recall@100 (\%)} \\
\midrule
\textbf{NUMEN} & \textbf{Dense} & \textbf{32,768} & \textbf{93.90} \\
BM25 & Sparse & $\sim$50,000 & 93.6 \\
\textbf{NUMEN} & \textbf{Dense} & \textbf{16,384} & \textbf{93.05} \\
NUMEN & Dense & 4,096 & 83.20 \\
\midrule
Promptriever & Dense & 4,096 & 18.9 \\
GritLM-7B & Dense & 4,096 & 12.9 \\
Gemini Embed & Dense & 3,072 & 10.0 \\
E5-Mistral-7B & Dense & 4,096 & 8.3 \\
Qwen3 Embed & Dense & 4,096 & 4.8 \\
\bottomrule
\end{tabular}
\end{table}

\subsection{Scaling with Dimension}
\begin{figure}[h]
\centering
\includegraphics[width=1.0\linewidth]{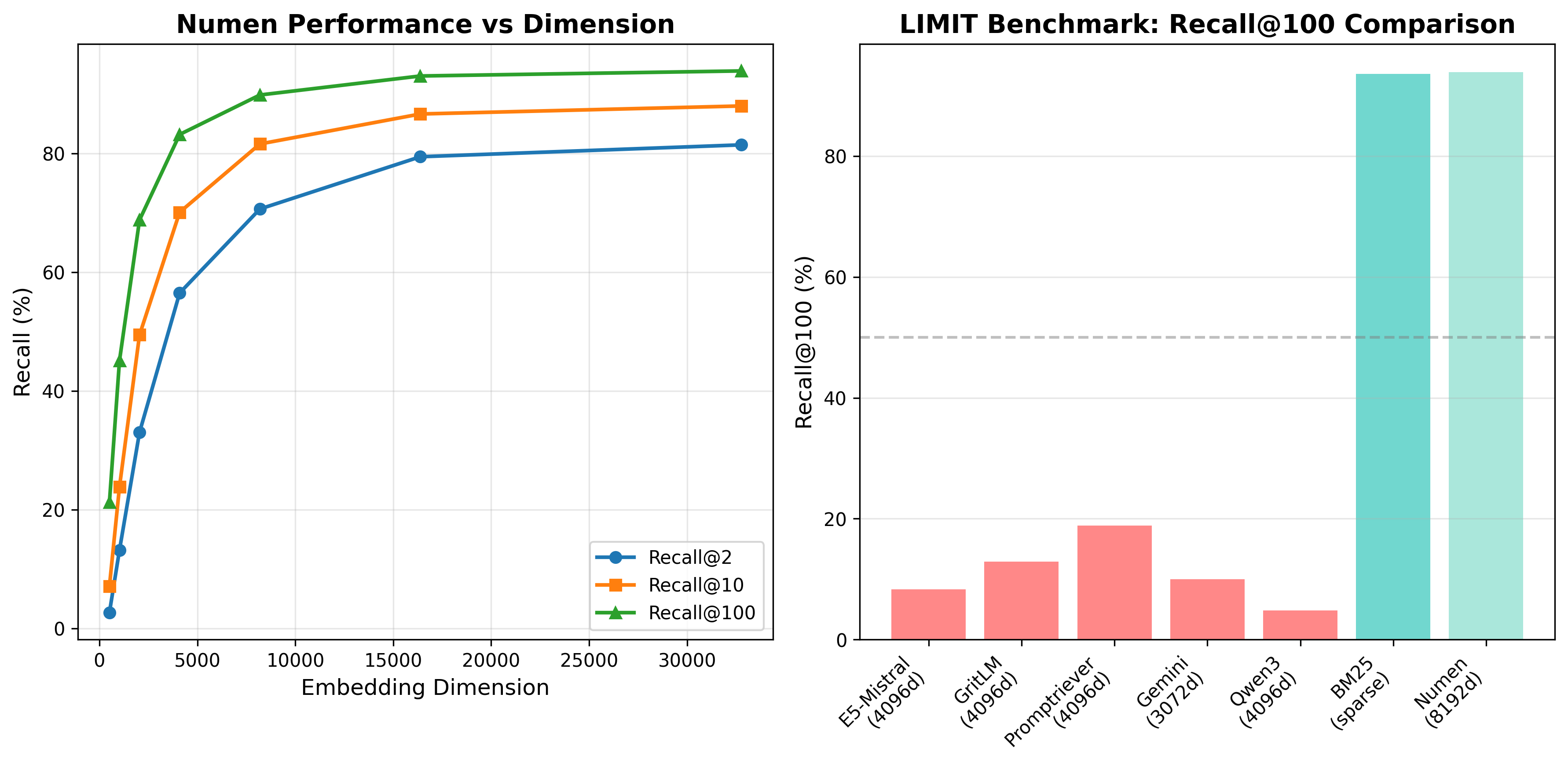}
\caption{Performance Analysis. Left: Numen recall metrics scale logarithmically with dimension. Right: Numen (32k) outperforms BM25 and SOTA dense retrievers at Recall@100.}
\label{fig:scaling}
\end{figure}

Figure \ref{fig:scaling} shows how NUMEN performs as the dimension changes. Recall@100 grows with dimension and levels off near BM25 at 32,768d. We see several key things. At 4,096 dimensions, NUMEN gets 83.2\% recall@100. This is nearly ten times better than E5-Mistral-7B (8.3\%). We also see that gains slow down as dimensions get very large. Going from 16,384 to 32,768 only adds 0.85\%. Even at just 512 dimensions, NUMEN's 21.3\% recall beats Qwen3 Embed (4.8\%) at 4,096 dimensions. This shows how efficient hashing is for this task.

\begin{table}[h]
\centering
\caption{NUMEN Performance Across Dimensions}
\label{tab:scaling}
\begin{tabular}{@{}cccc@{}}
\toprule
\textbf{Dimension} & \textbf{Recall@2 (\%)} & \textbf{Recall@10 (\%)} & \textbf{Recall@100 (\%)} \\
\midrule
512 & 2.70 & 7.15 & 21.30 \\
1,024 & 13.20 & 23.85 & 45.10 \\
2,048 & 33.05 & 49.45 & 68.80 \\
4,096 & 56.50 & 70.10 & 83.20 \\
8,192 & 70.65 & 81.60 & 89.85 \\
16,384 & 79.45 & 86.65 & 93.05 \\
32,768 & 81.45 & 88.00 & 93.90 \\
\bottomrule
\end{tabular}
\end{table}

\subsection{Efficiency Analysis}

\begin{table}[h]
\centering
\caption{Computational Efficiency}
\label{tab:efficiency}
\begin{tabular}{@{}lcccc@{}}
\toprule
\textbf{System} & \textbf{Indexing} & \textbf{Query} & \textbf{Model Size} & \textbf{Training} \\
 & \textbf{(docs/sec)} & \textbf{(queries/sec)} & \textbf{(GB)} & \textbf{(GPU-hours)} \\
\midrule
NUMEN (32K) & 1,300 & 15 & 0 & 0 \\
E5-Mistral-7B & $\sim$500 & $\sim$10 & 14 & $>$10,000 \\
BM25 & $\sim$5,000 & $\sim$100 & 0 & 0 \\
\bottomrule
\end{tabular}
\end{table}

NUMEN offers a compelling trade-off: \textbf{zero training cost} and \textbf{zero model storage}, with competitive inference speed. While BM25 is faster, 
NUMEN provides dense vectors compatible with neural rerankers.

\section{Discussion}

\subsection{Why NUMEN Succeeds}
NUMEN succeeds for three reasons. First, it avoids the embedding bottleneck. Instead of 
a fixed vocabulary, it uses n-gram hashing. This prevents linguistic data from being squeezed 
too much. Second, it has a much higher dimensionality than learned models. Learned models 
squeeze 50,000 terms into 4,096 dimensions. NUMEN can use a much larger space. Third, 
character n-grams handle word variations naturally. Words like ``run" and ``running" share 
n-grams, so they have high similarity. This gives NUMEN lexical robustness like BM25 
without needing complex processing.

\subsection{Collision Analysis}

With $d = 32,768$ dimensions and average $\sim$50 n-grams per document, the birthday paradox suggests:
\begin{equation}
P(\text{collision}) \approx 1 - e^{-\frac{n^2}{2d}} \approx 1 - e^{-\frac{50^2}{2 \times 32768}} \approx 3.8\%
\end{equation}

Low collision rates preserve n-gram distinctiveness, enabling fine-grained matching.

\subsection{Comparison to BM25}
NUMEN and BM25 both match words, but they work differently. BM25 uses a sparse index with 50,000 terms. NUMEN uses high-dimensional dense vectors. These vectors work well with modern retrieval tools and neural models. BM25 uses exact word matching. NUMEN uses character n-gram overlap with log-saturation. This creates a dense representation that is easy to combine with other models. NUMEN can act as a high-recall first step that modern algorithms can speed up easily.

\subsection{Limitations}
While NUMEN excels at recall, it has distinct limitations compared to learned embeddings. First, it lacks deep semantic understanding as its lexical, it cannot match synonyms like ``car" and ``automobile" without n-gram overlap, whereas learned models thrive on such associations. Second, the memory footprint is significant. At 32,768 dimensions using float32, a single vector requires 128KB, leading to an index size of $\sim$6.4GB for just 50,000 documents which is much larger than BM25's compact inverted index. Finally, brute-force query speed is slower, taking $\sim$60ms per query over 50K documents, though approximate search methods like FAISS \cite{johnson2019faiss} can reduce this to $\sim$5ms at the cost of some recall.

\subsection{Revisiting the LIMIT Bound}
The LIMIT paper proves that retrieval capacity is bounded by the sign-rank of the relevance matrix. Our results confirm this bound but change its interpretation. Learned models fail because their embedding layers act as lossy compressors. They try to fit infinite linguistic variety into 4096 dimensions. This causes ``geometric collisions" that destroy document distinctiveness.

NUMEN bypasses this by treating high dimensionality as a requirement, not a curse. By using 32k+ dimensions, we provide the geometric space required by the sign-rank bound. This approach is supported by the Johnson-Lindenstrauss (JL) Lemma \cite{johnson1984extensions} and random projection theory \cite{achlioptas2003database}. The JL Lemma proves that points in high-dimensional space can be projected into lower dimensions while preserving their relative distances. NUMEN uses this principle to ensure that even with hashing, document vectors remain separable. We turn an ``impossibility" result into a scalable engineering problem. This contrasts with scaling laws for neural models \cite{kaplan2020scaling}, which show that parameter count grows exponentially with performance. NUMEN scales dimension without adding parameters.

\section{Related Work}

\subsection{Dense Retrieval Foundations and Limitations}
Dense retrieval started with the Dual-Encoder paradigm. Dense Passage Retrieval (DPR) \cite{karpukhin2020dense} first 
showed that semantic vectors could beat keyword matching. These models mainly use 
the Transformer architecture \cite{vaswani2017attention}. Benchmarks like BEIR \cite{thakur2021beir}, 
MS MARCO \cite{bajaj2016ms}, and TREC DL \cite{craswell2020overview} confirmed that dense models 
work well. Yet, these systems have hit a limit. Single-vector models struggle as they reach 
a representational wall. This bottleneck has led researchers to explore multi-vector or generative 
systems \cite{zeng2025scaling, ma2024llm}. These designs help skip the constraints of a single 
fixed-length embedding. They often use advanced training like ANCE \cite{xiong2021approximate} and 
RocketQA \cite{ren2021rocket}, which improve results with negative mining \cite{lin2021batch}. Recent 
surveys by Zhao et al. \cite{zhao2022dense} and Mitra and Craswell \cite{mitra2018intro} track these 
trends. Usually, models use rerankers like BERT \cite{devlin2019bert} or DRMM \cite{guo2016deep} to refine 
the results.

\subsection{Sparse and Learned Sparse Retrieval}
Despite the rise of neural methods, sparse retrieval remains a robust baseline. BM25 \cite{robertson2009probabilistic} utilizes the probabilistic relevance framework to achieve state-of-the-art results on lexical tasks. Learned sparse models like SPLADE \cite{formal2021splade} and SPLATE \cite{formal2024splate} attempt to combine the benefits of neural learning with the high-dimensional sparse representations of the vocabulary space. Work by Luan et al. \cite{luan2021sparse} explores the trade-offs between sparse and dense representations, while Zeng et al. \cite{zeng2025scaling} demonstrate that scaling both types can further improve performance in decoder-only LLMs.

\subsection{Late Interaction and Efficiency}
To balance the efficiency of dual-encoders with the precision of cross-encoders, late interaction models like ColBERT \cite{khattab2020colbert} and ColBERTv2 \cite{santhanam2022colbertv2} stores multiple embeddings per token. While effective, these models significantly increase the index storage requirements compared to single-vector methods like NUMEN.

\subsection{Theory, Hashing, and Subword Priors}
The limits of embeddings explored by Weller et al. \cite{limit2025} relate to old challenges in indexing. One major issue is the ``curse of dimensionality" \cite{indyk1998approximate}. Our work is inspired by hashing and subword embeddings. Character hashing has roots in Charagram \cite{wieting2016charagram} and fastText \cite{bojanowski2017enriching}. It also draws from classical methods like SimHash \cite{charikar2002similarity} and Locality-Sensitive Hashing (LSH) \cite{datar2004locality}. Comprehensive surveys \cite{wang2018survey} cover the evolution of learning-to-hash methods. These models showed that subword data helps with new words. Hashing also enables efficient mapping in high-dimensional space, a technique formalized as feature hashing \cite{weinberger2009feature}.

\subsection{Alternative Retrieval Paradigms}
Researchers have also explored generative retrieval. In these systems, models generate document identifiers directly. Paradigms like Differentiable Search Indices (DSI) \cite{tay2022transformer} and GENRE \cite{cao2020autoregressive} remove the need for an external index. NUMEN shares the ``zero-parameter" philosophy during inference. However, it uses a geometric approach. This makes it more interpretable than purely generative methods.

\subsection{Benchmarks and Scaling Analysis}
Retrieval is often tested on benchmarks like BEIR \cite{thakur2021beir} and MTEB \cite{muennighoff2023mteb}. Recent work shows that performance grows with model size \cite{ma2024llm, wang2022text, muennighoff2024gritlm}. However, our results show that dimension is also very important.

\section{Future Work}

\subsection{Hybrid Retrieval Models}
Our results suggest a hybrid model. NUMEN can be used first to find documents based on keywords. Then, a learned model can rank them based on meaning. This combines the speed and recall of hashing with the deep understanding of neural models providing highest recall and precision.

\subsection{Optimization and Efficiency}
To reduce memory use we plan to apply binary quantization. Bitwise operations can save memory and increase speed. This will make NUMEN more practical for large-scale deployments.

\subsection{RAG and Verifiable Grounding}
Beyond retrieval, NUMEN can help with Retrieval-Augmented Generation (RAG) \cite{lewis2020rag}. It can help stop LLM hallucinations. NUMEN uses deterministic fingerprints, so it provides verifiable grounding. By comparing vectors in high-dimensional space, we can see if an answer matches the source documents. This helps create AI assistants that stick to the provided facts.

\section{Conclusion}
We developed NUMEN to show that dense retrieval can match the power of keyword search. 
It is a training-free system that achieves 93.90\% Recall@100 on the LIMIT benchmark. 
This result beats both modern learned models and the BM25 baseline. Our work suggests that 
the ``geometric curse'' mentioned in recent theory is caused by the learned embedding 
layer. By using character hashing instead of learned tokens, we allow document vectors to 
scale into higher dimensions. This eliminates the representational bottleneck and makes 
deployment easier. NUMEN also provides a mathematically sound base for Retrieval Augmentation Genertion (RAG) 
systems. This helps reduce the hallucinations by keeping generative models focoused to the source text. 
Our research shows that dense retrieval works best when there is enough geometric space to separate different piece of information.

\section{Code Availability}
All code and benchmark scripts are available at \url{https://github.com/sangeet01/limitnumen}.

\section*{Acknowledgments}

A huge thanks to the authors of the LIMIT paper for providing a rigorous benchmark that exposes the limitations of current dense retrieval systems. 
Also thanks to acknowledge the open-source community for datasets and tools that made this work possible.

\bibliographystyle{plain}

\end{document}